\documentclass[12pt]{iopart}
\usepackage{graphicx}
\usepackage{dcolumn}
\usepackage{bm}
\usepackage{color}
\usepackage{todonotes}
\newcommand{\Gvec}[1]{\boldsymbol{#1}}
\renewcommand{\vec}[1]{\mathbf{#1}}
\newcommand{\cP}{\mathcal{P}}

\renewcommand{\div}{\nabla\cdot}
\usepackage{tikz}

\newcommand{\eqref}[1]{(\ref{#1})}

\usepackage[pdftex]{hyperref}
\usetikzlibrary{patterns}
\begin{document}

\title{Sedimentation of self-propelled Janus colloids: polarization and pressure}

\author{Felix Ginot}
\address{Ceramic Synthesis and Functionalization Lab, UMR3080 CNRS-Saint-Gobain, 84306 Cavaillon, France}
\address{Univ. Lyon, Universit\'e Claude Bernard Lyon 1, CNRS,UMR 5306, Institut Lumi\`ere Mati\`ere, F-69622 Villeurbanne, France.}

\author{Alexandre Solon}
\address{Department of Physics, Massachusetts Institute of Technology, Cambridge, Massachusetts 02139, USA}

\author{Yariv Kafri}
\address{Department of Physics, Technion, Haifa, 32000, Israel}

\author{Christophe Ybert}
\address{Univ. Lyon, Universit\'e Claude Bernard Lyon 1, CNRS,UMR 5306, Institut Lumi\`ere Mati\`ere, F-69622 Villeurbanne, France.}

\author{Julien Tailleur}
\address{Universit\'e Paris Diderot, Sorbonne Paris Cit\'e, MSC, UMR 7057 CNRS, 75205 Paris, France}

\author{Cecile Cottin-Bizonne}
\address{Univ. Lyon, Universit\'e Claude Bernard Lyon 1, CNRS,UMR 5306, Institut Lumi\`ere Mati\`ere, F-69622 Villeurbanne, France.}

\date{\today}

\begin{abstract}
We study experimentally---using Janus colloids---and theoretically--- using Active Brownian Particles---the sedimentation of dilute active colloids. We first confirm the existence of an exponential density profile. We show experimentally the emergence of a polarized steady state outside the effective equilibrium regime, \textit{i.e.} when $v_s$ is not much smaller than the propulsion speed $v_0$. The experimental distribution of polarization is very well described by the theoretical prediction with no fitting parameter. We then discuss and compare three different definitions of pressure for sedimenting particles: the weight of particles above a given height, the flux of momentum and active impulse, and the force density measured by pressure gauges.
\end{abstract}
\maketitle

\section*{Introduction}
\label{sec:intro}

The sedimentation of active particles has recently attracted a lot of interest, both  experimentally~\cite{Palacci2010,Ginot2015} and theoretically~\cite{Tailleur2009a,Nash:2010,enculescu_active_2011,Wolff2013EPJE,Szamel2014,Solon2015EPJST,stark2016EPJST,vachier2017}. In the simplest of limits in which the interactions between particles can be neglected and their sedimentation speed $v_s$ is much smaller than their self-propulsion speed $v_0$, the system behaves as an equilibrium one, leading to an exponential density profile
\begin{equation}
\rho(z) \propto \exp(- mg z/kT_{\rm eff}).
\end{equation}
The effective temperature is then given by a Stokes-Einstein relation $k T_{\rm eff} \equiv D/\mu$, with $D$ and $\mu$ the diffusivity and the mobility of the particles. This regime was observed experimentally for self-diffusiophoretic Janus colloids~\cite{Palacci2010}. The impact of interactions between particles on the above small $v_s/v_0$ regime was recently explored experimentally and numerically in~\cite{Ginot2015}. 

In this article, we consider experimentally and theoretically the fate of dilute active sedimenting systems when the sedimentation speed cannot be neglected, i.e. beyond the effective equilibrium regime. We use self-propelled Janus colloids in 2D as a model experimental active system, and model them using non-interacting Active Brownian Particles (ABPs, see Section~\ref{sec:manip}). We first consider the distribution of particles in the sedimentation profile (Section~\ref{sec:sedim}). Our experiments show that the gravity field leads to a polarized steady state in agreement with earlier theoretical predictions~\cite{enculescu_active_2011,Solon2015EPJST,wagner2017JSM}. Furthermore, the distribution of orientations of the particles within the exponential sedimentation profile agrees quantitatively, without any fitting parameter, with the one predicted analytically for sedimenting active Brownian particles~\cite{Solon2015EPJST}.

The pressure of active particles has attracted a lot of interest recently~\cite{mallory2014anomalous,yang2014aggregation,takatori2014swim,Solon2015natphys,solonPressure2015b,basu2015statistical,winkler2015virial,smallenburg2015swim,speck2016ideal,joyeux2016pressure,falasco2016mesoscopic,razin2017generalized,junot2017active,Fily2018JPA}. We discuss   the definition of pressure for sedimenting active particles~\cite{Ginot2015} (Section~\ref{sec:Pressure}). For our ABP  model, we give a clear interpretation of a bulk pressure, defined as the weight exerted on the system above a certain height~\cite{Barrat1992, Biben1993, Piazza1993,Piazza2007PRL,Piazza2012,Ginot2015}, in terms of momentum transfer. We give a complete characterization of the latter in terms of correlators measured in the bulk of the system and a recently introduced active impulse~\cite{Fily2018JPA}. Excellent agreement is shown between experimental measurements of the pressure and these bulk observables. Finally, we discuss whether such a definition of pressure can be related to force densities exerted on pressure gauges.

\section{Experimental setup and theoretical model.}
\label{sec:manip}
\begin{figure}
	\centering
   	\includegraphics[width=0.9\columnwidth]{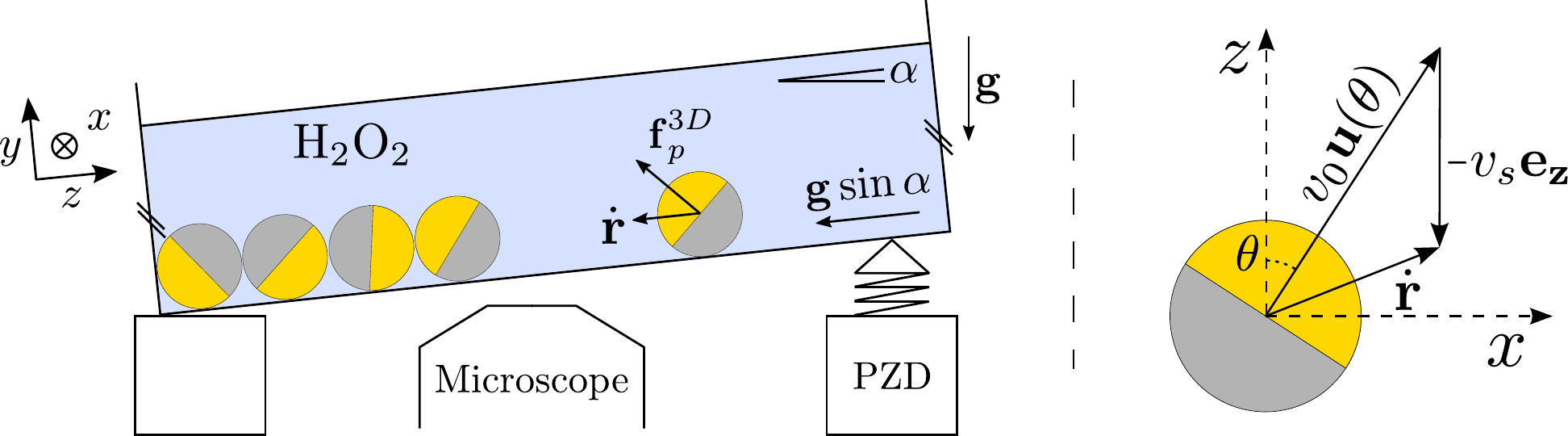}
	\caption{{\bf Left:} Illustration of the experimental setup. Janus particles are immersed in a hydrogen peroxide bath and sediment towards the bottom of the experimental cell. They then form a monolayer where each particle moves  with 2D velocity $\dot{\vec r}$ resulting from both self-propulsion and sedimentation.
 Using a piezoelectric device (PZD) we  tilt the experimental cell by an angle $\alpha$ which leads to a reduced gravity field $\vec g \sin \alpha$. {\bf Right:} 2D scheme of a Janus particle under sedimentation. The particle moves forward with a mean velocity $v_0$ and an orientation $\theta$, but due to the sedimentation velocity $v_s$, the (average) velocity vector is $\dot \vec r =v_0 \vec u (\theta) - v_s\vec e_z$.}
	\label{fig:sketch}
\end{figure}

When immersed in a hydrogen peroxide ($\textrm{H}_2\textrm{O}_2$) bath, gold Janus colloids of radius $a=1.1 \pm 0.1 \mu$m half coated with Platinum become active and propel themselves with a force ${\bf  f}_{p}^{3D}$ by self-diffusiophoresis~\cite{Paxton_jacs-2004}.
Their density being very high around $\rho = 11$ g.cm$^{-3}$, the colloids immediately sediment onto the flat bottom of an experimental cell to form a bidimensional layer of particles. 
Due to the huge reservoir of peroxide, activity can be considered as constant during each experiment.
For each experiments we record movies of 5000 images @ 20 fps, for a total duration of 250s. This experimental set-up, sketched in figure~\ref{fig:sketch}, was previously used to study the cluster phase~\cite{Ginot2018} and the weak sedimentation limit~\cite{Ginot2015} of active Janus colloids.

Here, using a piezoelectric module, we tilt the experimental cell with an angle $\alpha$ to create a reduced gravity field $\vec g\sin \alpha$, leading to a controllable sedimentation velocity $v_s$, that we take along the $z$-axis: $-v_s \vec e_z$. 
We start with an angle $\alpha$ as small as possible, 
and we increase $v_s$ by increasing $\alpha$, until all the gas collapses into the dense phase.
In the following we denote $v_0\vec u(\theta)$ the 2D propulsion velocity of a colloid in the ($x,z$) plane (See Fig~\ref{fig:sketch}).
Note that, experimentally, we measure the velocity of a particle at position $\vec r$: $\dot {\vec r}=v_0\vec u(\theta)-v_s \vec e_z$. For each experiment we measure $v_0$, which is found around $4 \pm 0.2 \mu $m.s$^{-1}$, and $v_s$ (see \ref{sec:vs} and \ref{sec:3dproj} for details). In the following, we will present experimental results for different realizations corresponding to values of the ratio $v_s/v_0$ from $\sim 0.08$ to $\sim 0.28$.

To account for our experimental results, we model the colloids as active Brownian particles~\cite{Fily2012} that are self-propelled at a constant velocity $v_0$ along their internal direction of motion $\vec u$ and subject to rotational diffusion.
As in the experiments, the motion of the particles is restricted to the 2D plane parallel to the bottom plate and subject to a sedimentation velocity $-v_s\vec e_z$ downward along the $z$-axis. 
Furthermore, we will assume that the orientation vectors $\vec u$ of the particles are also restricted to this 2D plane. This is not the case experimentally since the propulsion force ${\bf f}_p^{3D}$ can point in any direction in 3D. However, we do not have experimental access to the 3D statistics of ${\bf f}_p^{3D}$ and we show in~\ref{sec:3dproj} that allowing rotational diffusion in 3D for the ABPs lead only to quantitatively similar results with small corrections, so that our experimental data are not able to distinguish between the two situations. 
For simplicity, we thus consider a propulsion velocity of fixed norm and 2D orientation vector $\vec u(\theta)=(-\sin\theta,\cos\theta)$, subject to rotational diffusion with diffusion coefficient $D_r$. 
The overdamped dynamics of a particle at position $\vec r$ and with a sedimentation speed $v_s$ then follows the Langevin equation
\begin{equation}
  \label{eq:LangevinABP}
  \dot {\vec r}=v_0\vec u(\theta)-v_s \vec e_z;\qquad \dot\theta=\sqrt{2D_r}\xi\;,
\end{equation}
where $\xi$ is a Gaussian white noise with zero mean and unit variance $\langle \xi(t)\xi(t')\rangle = \delta(t-t')$.
Eq.~(\ref{eq:LangevinABP}) does not include interactions between particles and thus only attempts to describe the dilute gaseous phase of the experiments. 
The persistence time $\tau_p \equiv D_r^{-1}$ and length $l_p \equiv v_0 D_r^{-1}$ provide natural time and length units.
Numerically, we integrate Eq.~(\ref{eq:LangevinABP}) using Euler time-stepping with time step $dt=0.1\tau_p$.

\section{Sedimentation profile and polarization}
\label{sec:sedim}
The steady state distribution of sedimenting ABPs described by Eq.~(\ref{eq:LangevinABP}) is an exponential density profile~\cite{enculescu_active_2011,Solon2015EPJST,stark2016EPJST,wagner2017JSM}.
Indeed, the Fokker-Planck equation for the probability $\cP(\vec r,\theta,t)$ to find a particle at position $\vec r$ with orientation $\theta$ at time $t$ reads
\begin{equation}
  \label{eq:FP}
  \partial_t\cP=-\div \left[(v_0\vec u-v_s\vec e_z)\cP\right]+D_r\partial_\theta^2\cP\;,
\end{equation}
and by symmetry $\cP(\vec r,\theta)=\cP(z,\theta)$.  
As shown in Ref~\cite{Solon2015EPJST}, Eq.~(\ref{eq:FP}) can be solved by separation of variables. 
Writing $P(z,\theta)=f(\theta)\rho(z)$ in Eq.~(\ref{eq:FP}), $\rho$ and $f$ satisfy in steady state

  \begin{equation}
  \rho'(z)=-\rho(z)/\lambda\;, 
    \label{eq:FP-rho}
    \end{equation}
  \label{eq:FP-f}

  \begin{equation}
  f''(\theta)=-\frac{1}{D_r\lambda}(v_0\cos\theta -v_s)f(\theta)=0\;.
  \label{eq:FP-rho}
  \end{equation}

The density profile is thus of the form $\rho(z)\propto e^{-z/\lambda}$, with a sedimentation length $\lambda$. 
The solutions of Eq.~(\ref{eq:FP-rho}) are Mathieu functions. 
The periodicity of $f(\theta)$ then implies~\cite{Solon2015EPJST} $-4 v_s/(\lambda D_r)=a_0(-2v_0/(\lambda D_r))$ where $a_0$ is the first characteristic value of the Mathieu equation~(see~\cite{alhargan1996complete} for properties of the solutions of Eq.~(\ref{eq:FP-rho})). 
Expanding $a_0$ for small $v_s/v_0$~\cite{alhargan1996complete}, one gets the sedimentation length $\lambda$ as:
\begin{equation}
  \label{eq:lambda-expansion}
  \lambda=\frac{v_0^2}{2D_r v_s}\left[1-\frac{7}{4}\left(\frac{v_s}{v_0}\right)^2+O\left(\left(\frac{v_s}{v_0}\right)^4\right)\right].
\end{equation}
At the same order in $v_s/v_0$, one gets for the orientation distribution:
\begin{equation}
  \label{eq:ftheta2dABPs}
  2\pi f(\theta)=1+\frac{2v_s}{v_0}\cos\theta+\frac{v_s^2}{2v_0^2}\cos 2\theta+O\left(\left(\frac{v_s}{v_0}\right)^3\right).
\end{equation}

\begin{figure}
	\centering
  	\includegraphics[width=0.7\columnwidth]{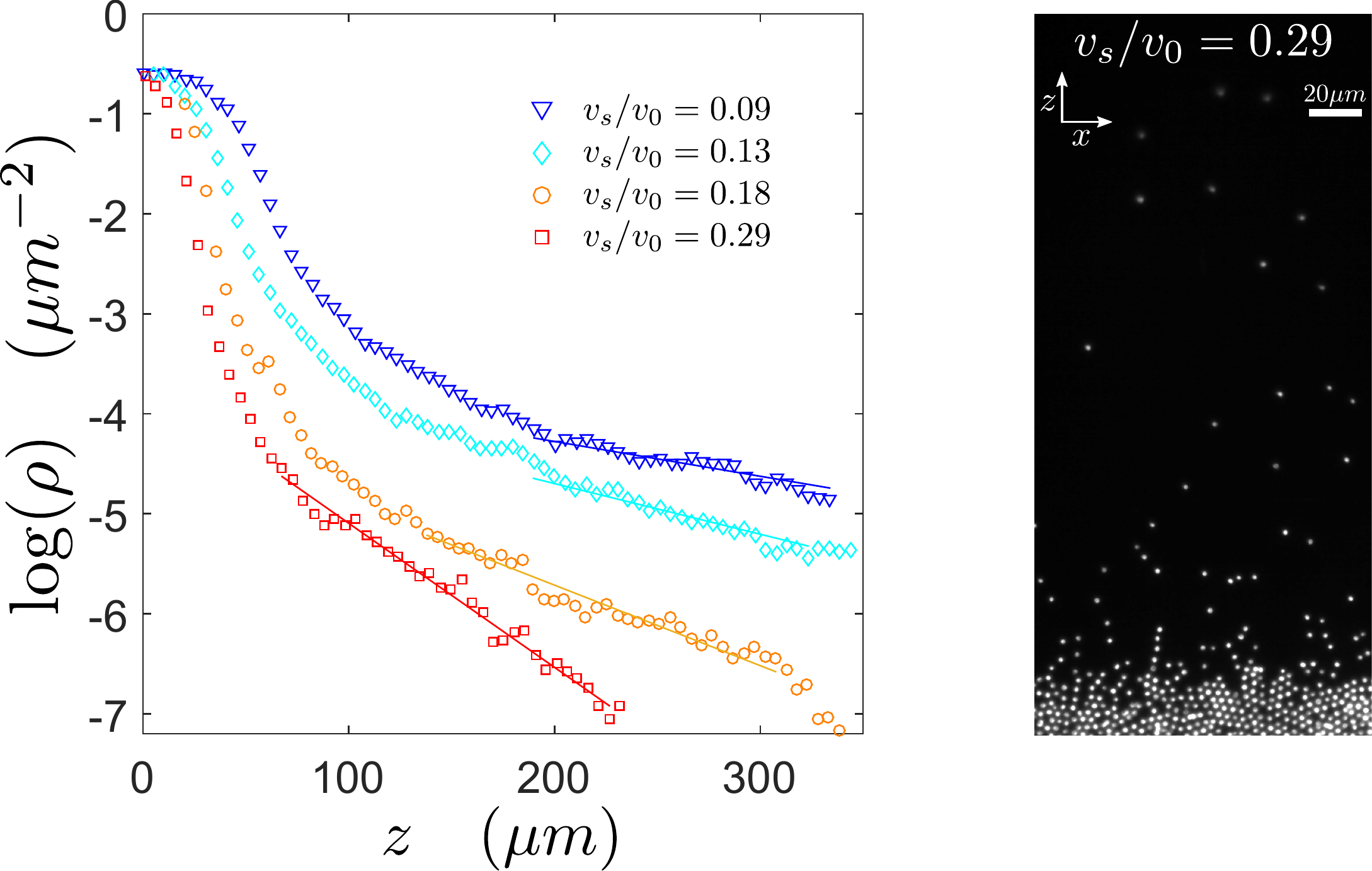}
	\caption{{\bf Left:} $\rho(z)$ for $v_s/v_0 = 0.09, 0.13, 0.18, 0.29$. We observe an exponential decay in the dilute gas phase, where particle interactions become negligible. Note that for a better display the origin of $z$ is arbitrary. 
 The plain lines correspond to a joint fit of all data sets with one free parameter, $D_r$, which is found to be  $D_r=0.08 \pm 0.003 s^{-1}$. {\bf Right:} Picture of the experimental system for $v_s/v_0 = 0.29$. Blurry particles at the top are due to the defocus induced by the tilt of the sample.}
	\label{fig:rho_de_Z}
\end{figure}

\paragraph*{Density profile.}
We first measure experimentally the density profile $\rho(z)$.
To do so, we use a coarse graining operation ($5.1\mu m$ height slices) and both a spatial average over the x-axis and a time average over the experiment duration.
In Fig.~\ref{fig:rho_de_Z} the resulting experimental sedimentation profiles is shown for several values of $v_s/v_0$. The density profiles indeed exhibit an exponential decay in the dilute phase---which corresponds to a linear dependence in our semi-log plot---with a sedimentation length $\lambda$.

The theoretical expression for the sedimentation length Eq.~\eqref{eq:lambda-expansion} depends only on parameters that should be accessible experimentally. However, measuring precisely the rotational diffusion coefficient $D_r$ of the active colloids is difficult. The statistics of the velocity autocorrelation in the dilute gaz region is indeed too limited to accurately measure  $D_r$.
Therefore, we use $D_r$ as a free parameter and fit the density profiles shown in Fig~\ref{fig:rho_de_Z} with Eq.\eqref{eq:lambda-expansion}. This leads to $D_r=0.08 \pm 0.003 s^{-1}$. Note that this value is compatible with the Brownian estimate $D_r = 0.12 \pm 0.04 s^{-1}$ (the significant errorbar is due to 10\% polydispersity). Our slightly smaller measurement could be due to the proximity of the bottom surface.

\begin{figure}
	\centering
   	\includegraphics[width=0.7\columnwidth]{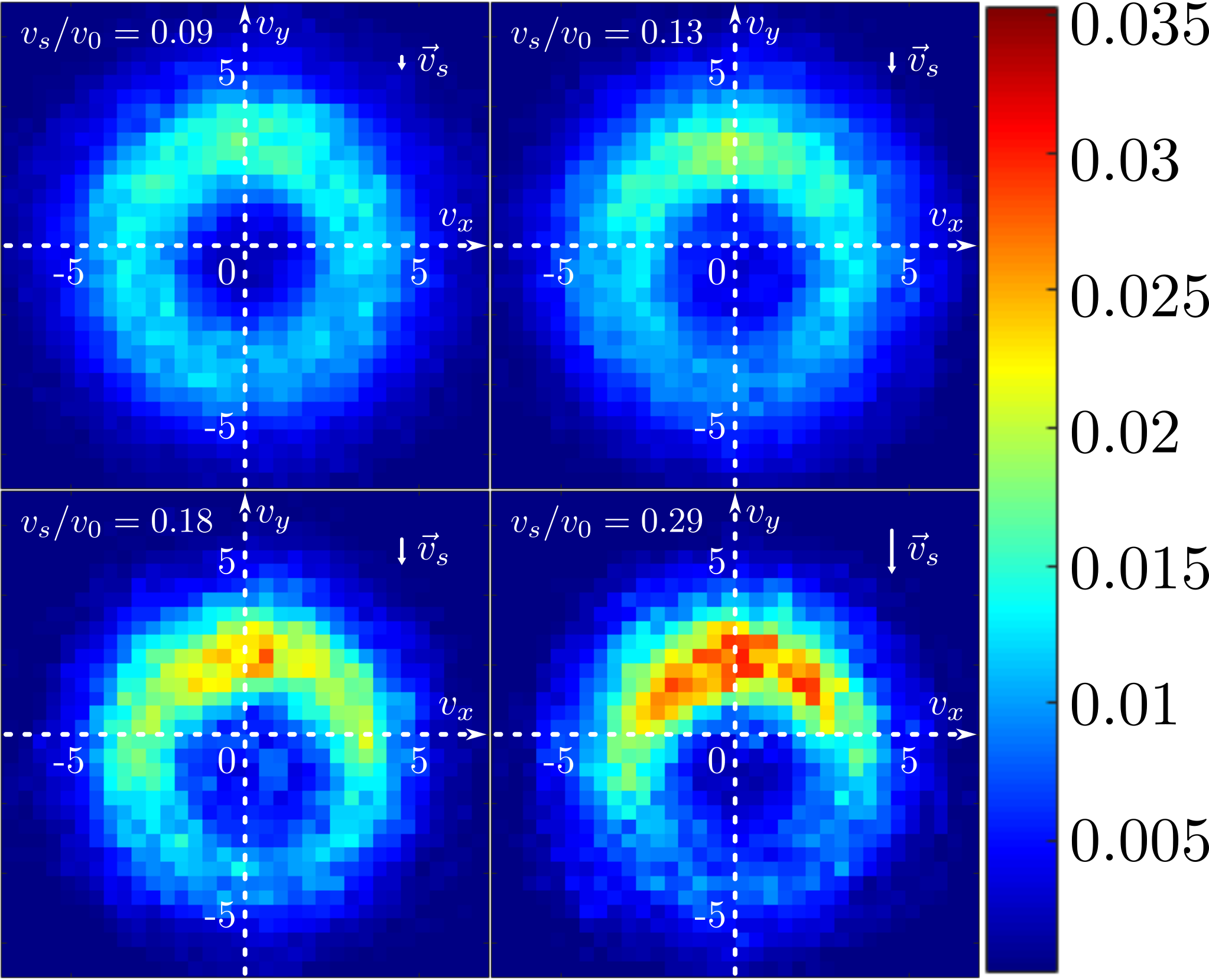}
	\caption{Probability distribution of the measured velocities $\dot {\bf r}$. The color code corresponds to a normalized distribution $P(\dot x, \dot z)$. The polarization becomes more apparent as $v_s/v_0$ increases. The downward offset of the distribution gives access to the sedimentation velocity $v_s$ (see \ref{sec:vs}), represented to scale by the white arrows in the upper right corner of each subfigure.}
	\label{fig:V2D}
\end{figure}

\paragraph*{Polarization.}
A remarkable feature of the sedimentation profile of active particles is the existence of a non-vanishing mean polarization. A first visualization of this polarization can be obtained from the distribution of velocities in the sedimentation profile.
We measure the instantaneous velocities of free particles $\dot {\bf r}$---positions are smoothed with a gaussian average over $1s$, particles are considered free if they have no neighbor in a radius of $5.1\mu$m---and build the corresponding 2D probability distribution function (see Fig.~\ref{fig:V2D}).
Note that $\dot {\bf r}$ corresponds to the observed velocity, which includes both the propulsion velocity $v_0 {\bf u}(\theta)$ and the sedimentation velocity $-v_s {\bf  e}_z$.

As expected, when the sedimentation field is negligible, for example when $v_s/v_0 \sim 0.09$, the distribution of velocities has an isotropic `ringlike' shape, with a radius~$\sim v_0$.
On the contrary, when the sedimentation speed increases, a striking behavior emerges as the microswimmers polarize against the gravity field: the distribution of velocities is no longer isotropic and colloids are most likely oriented upwards, leading to a strong peak of probability at the top of the ringlike distribution.
Note that, due to the sedimentation velocity, there is also a downward shift of the  center of the ring, which is clearly visible when $v_s/v_0$ is large enough.

\begin{figure}
	\centering
      	\includegraphics[width=0.5\columnwidth]{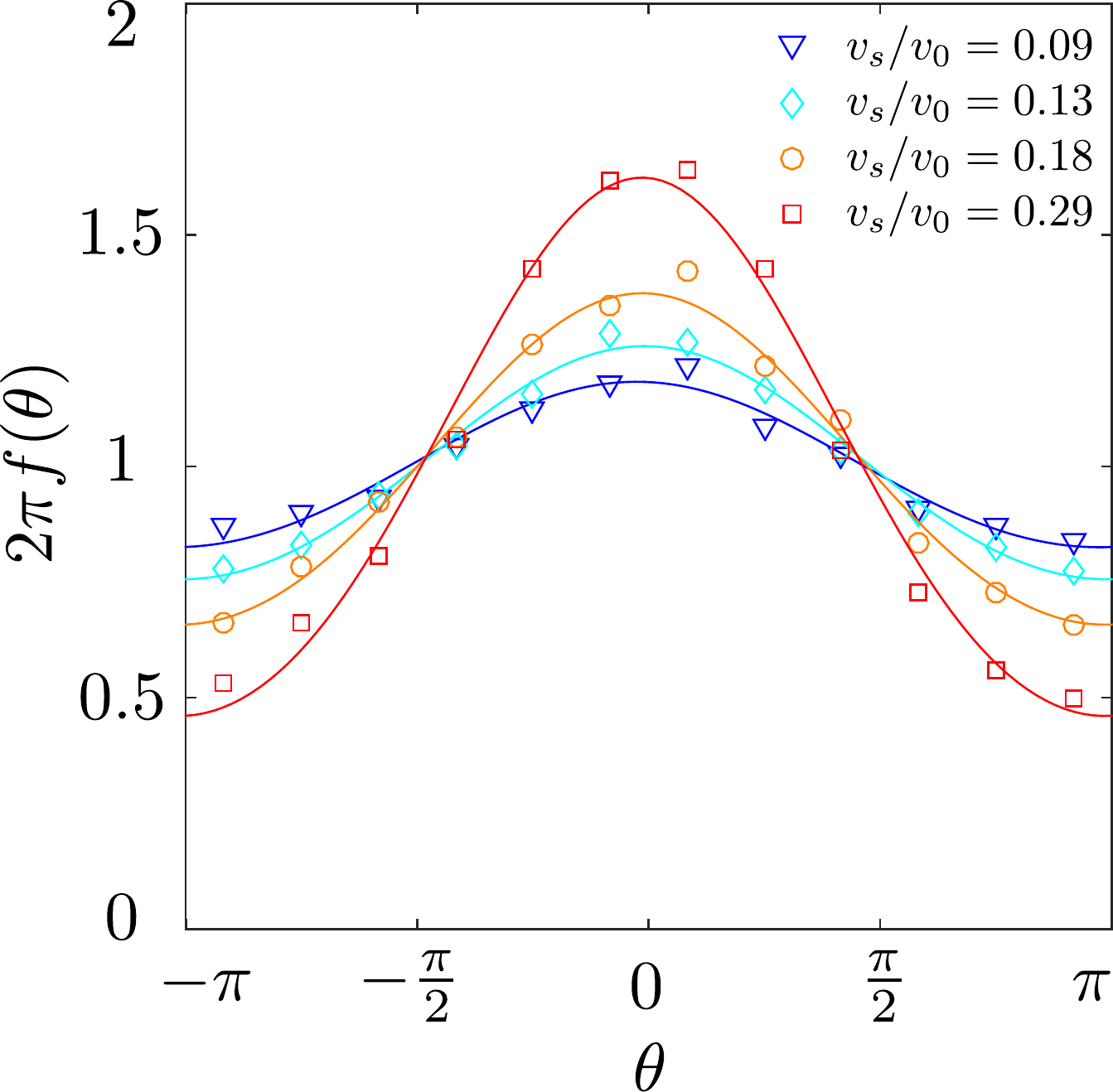}
	\caption{Distribution of particle orientations for increasing $v_s/v_0 = 0.09, 0.13, 0.18, 0.29$. Symbols correspond to experiments and solid lines to the theoretical predictions from Eq.~\eqref{eq:ftheta2dABPs}.}
	\label{fig:Pdf_theta}
\end{figure}

To compare with theoretical predictions based on the ABP model~\eqref{eq:LangevinABP}, we extract the orientation distribution $f(\theta)$ from the  experimental 2D probability distribution of $\dot {\bf r}$.
We plot the orientation distribution $2\pi f(\theta)$ in Fig.~\ref{fig:Pdf_theta} (symbols) against the theoretical prediction from Eq.~\eqref{eq:ftheta2dABPs} (solid lines).
Note that the agreement between experiments and theory is remarkable, without any fitting parameter. Modeling our self-propelled Janus colloids as active Brownian particles thus allows us to quantitatively account for their sedimentation profile.

\section{Pressures in sedimentation profiles}
\label{sec:Pressure}

Studying the pressure of active systems is a fascinating challenge for at least two reasons: first as the out-of-equilibrium fate of a thermodynamic state variable that controls phase equilibria and flows in passive systems. Then as a measurement of the force that active particles collectively exert on their environment. Since active particles exchange momentum with the environment, their momentum does not satisfy a conservation equation so that pressure cannot be unambiguously defined from their momentum flux. In particular, such a bulk definition is not, in general, equivalent to the force density exerted by active particles on a confining boundary~\cite{Solon2015natphys}.

There are notable exceptions to this lack of equation of state, such as  non-interacting Active Brownian Particles~\cite{mallory2014anomalous,yang2014aggregation,takatori2014swim,solonPressure2015b}. For such models, a homogeneous isotropic system exerts a force density on a container that can be expressed as observables measured in the bulk of the system, despite wall-dependent boundary layers. The question as to whether this extends to our polarized sedimentation profile is completely open. In section~\ref{sec:active_impulse}, we first show that a pressure defined as the weight exerted on the active system above a certain height can be related to momentum transfer in the bulk of the system, as suggested in~\cite{Ginot2015}. Nevertheless, we show in section~\ref{sec:mechanical_forces} that this bulk pressure cannot be related to the force density measured by a confining interface: it cannot be read with a pressure gauge. The difference between such a mechanical measurement and our bulk pressure however vanishes as $v_s/v_0 \to 0$, provided the pressure gauge is oriented orthogonally to the gravity field.

\subsection{Pressure as momentum flux: the active impulse}
\label{sec:active_impulse}

\begin{figure}
	\centering
  	\includegraphics[width=0.7\columnwidth]{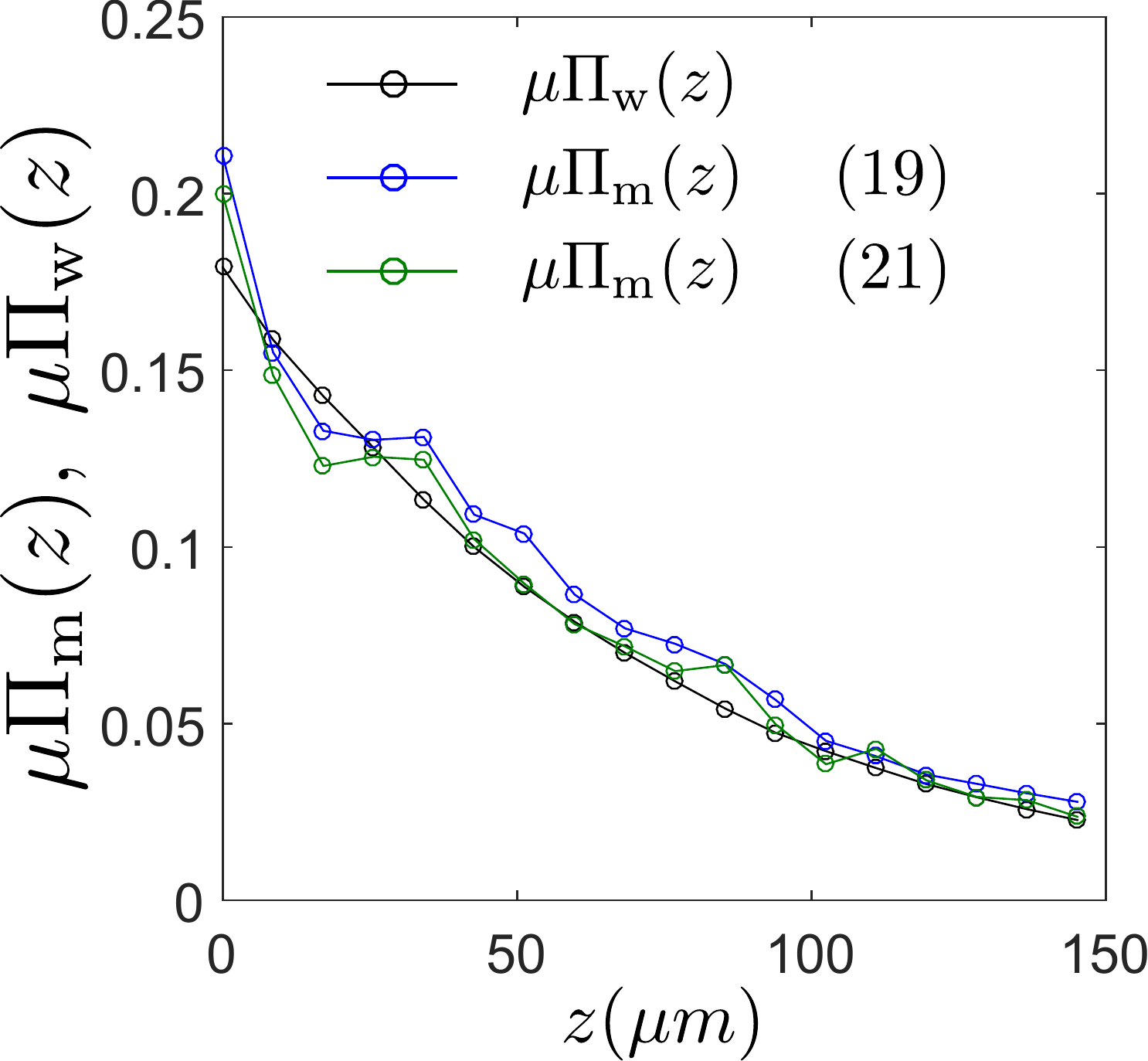}
	\caption{Pressure profiles for $v_s/v_0=0.29$. 
    In black, the pressure $\Pi_{\rm w}(z)$ measured experimentally by integrating the density above a height $z$, following Eq.\eqref{eq:P_experimental}. In blue, the effective momentum flux $\Pi_{\rm m}(z)$ predicted by Eq.~\eqref{eq:pisfinalMT} and, in green, its approximation Eq.~\eqref{eq:pi21}. Note that the curves for $\Pi_{\rm m}$ correspond to local measurements whereas $\Pi_{\rm w}(z)$ results from an integration. The noise level in the latter is thus much lower than in the former.}
	\label{fig:PI}
\end{figure}

\begin{figure}
	\centering
  	\includegraphics[width=0.7\columnwidth]{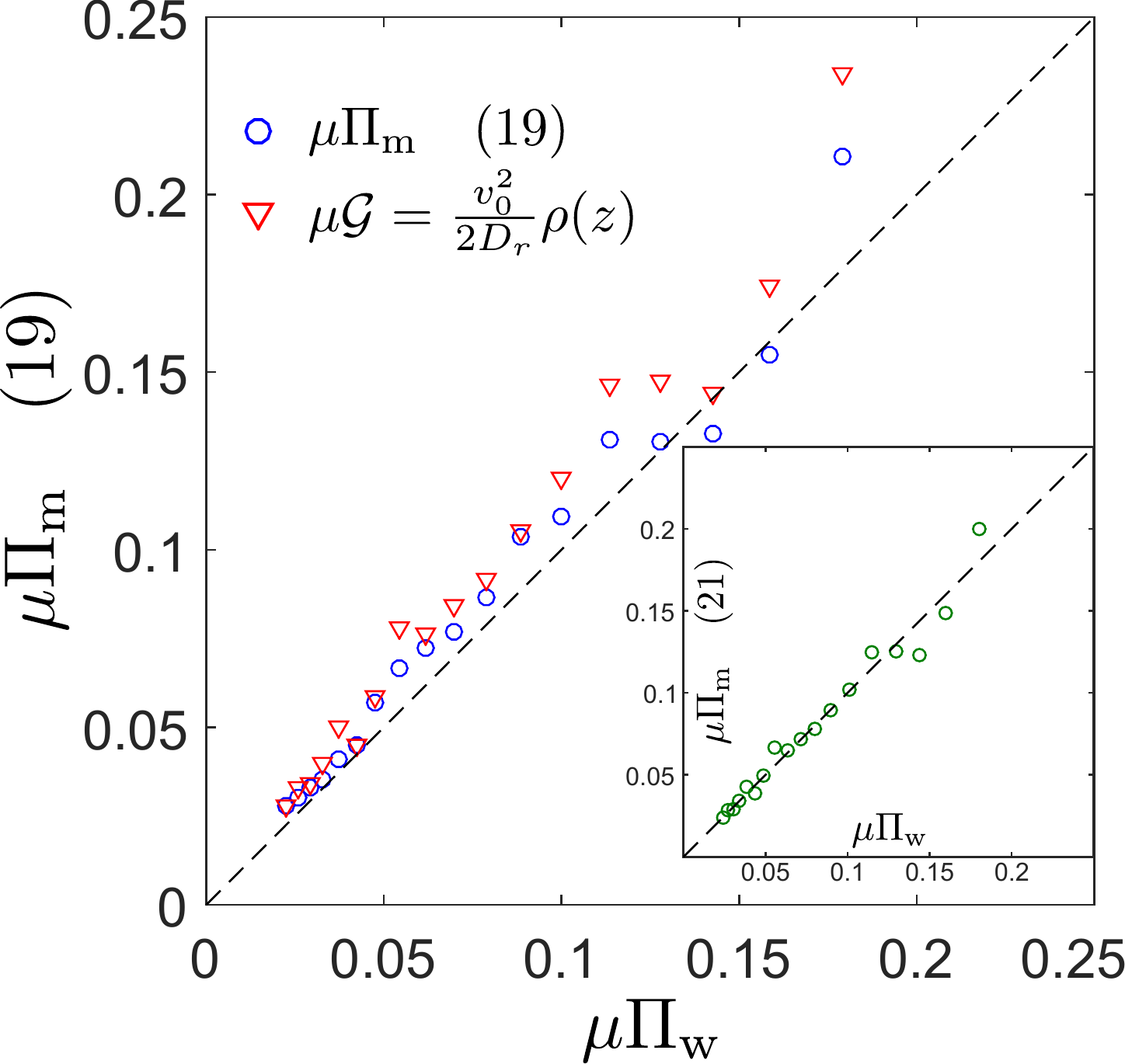}
	\caption{Parametric plot of $\mu\Pi_{\rm m}(z)$, Eq.~\eqref{eq:pisfinalMT}, vs $\mu\Pi_{\rm w}(z)$, Eq.~\eqref{eq:P_experimental}, for $v_s/v_0=0.29$ (blue symbols).    The black dotted line $y=x$ is a guide for the eye that corresponds to our theoretical predictions. A best affine fit $y=ax+b$ gives $a=1.04$ and $b=0.003$. The role of polarization for this value of $v_s/v_0$ can be visualized using $\mu{\cal G}=\frac{v_0^2}{2 D_r}\rho(z) $ instead of $\Pi_{\rm m}$ (red symbols), which leads to $a=1.16$ and $b=0.001$. The $1.16$ prefactor corresponds to the prediction~\eqref{eq:pi21}: $\left(1-\frac{7v_s^2}{4v_0^2}\right)^{-1}\simeq 1.17$. {\bf Inset}, parametric plot of the approximation~\eqref{eq:pi21} of $\mu\Pi_{\rm m}(z)$ vs $\mu\Pi_{\rm w}(z)$, which leads to $a=0.99$ and $b=0$.} 
	\label{fig:PI_19}
\end{figure}

In equilibrium, the equation of state relating the osmotic pressure to bulk properties of a system can be directly measured using a sedimentation profile~\cite{Barrat1992,Piazza2007PRL}, within a local density approximation. The underlying idea is that the total weight exerted on the particles above a given height $z$ is balanced by the osmotic pressure at this height so that the pressure can be measured as:
\begin{equation}\label{eq:psedim}
\Pi_{\rm w}(z)=\int_z^\infty m g_{\rm eff} \rho(z') dz'
\end{equation}
where $mg_{\rm eff}$ is the effective weight of the  particles.  Within the local density approximation, one then infers $\Pi_{\rm w}(\rho(z))$ from $\Pi_{\rm w}(z)$. Equation~\eqref{eq:psedim} can be seen as a consequence of momentum conservation above a height $z$: the incoming momentum flux, which we write $\Pi_{\rm m}(z)$, is balanced by the total external force (density) applied above a plane $z$, $\Pi_{\rm w}(z)$, which is the only external source of momentum. 

It is natural to ask whether this construction also applies to active systems, as recently proposed~\cite{Ginot2015}. Indeed, each active particle injects momentum into the active fluid so that momentum is not conserved.  This injection of momentum can be quantified using the active impulse $\Delta \vec p^{\rm a}$~\cite{Fily2018JPA}. For particle $i$, at position $\bf r_i$ and orientation $\theta_i$, $\Delta \vec p^{\rm a}_i$ measures the total momentum the particle will receive on average from the environment in the future:
\begin{equation}
\label{eq:activeimpulseABP}
\Delta \vec p^{\rm a}_i(t)=\int_t^\infty ds \, \overline{{\vec f}^{\rm p}_i(s)} \;,
\end{equation}
where $\vec f^{\rm p}_i=f^{\rm p} \vec u(\theta_i)$ is the propulsion force of the particle and the overbar represents an average over future histories, for fixed ${\bf r}_i(t)$ and $\theta_i(t)$. 
Interestingly, it was recently shown that a class of active systems, to which our ABP model~\eqref{eq:LangevinABP} belongs, admits a generalized conservation law: the sum of the particles' momenta and of their active impulses \cite{Fily2018JPA} form a conserved quantity. We now build upon this and show that, in this context, $\Pi_{\rm w}(z)$, defined in equation~\eqref{eq:psedim}, can be related to the flux of momentum and active impulse through an interface at height $z$.

We consider particles evolving under the dynamics~\eqref{eq:LangevinABP}. 
The dynamics of the microscopic density field $\hat \rho=\sum_i \delta(\vec r- \vec r_i)$ is given by
\begin{eqnarray}
\dot {\hat \rho} &=& - \nabla \cdot \hat {\vec J}\\
\hat {\vec J} &\equiv& - v_s \vec e_z \hat \rho + \sum_i v_0 \vec u(\theta_i) \delta(\vec r -\vec r_i)\;.
\end{eqnarray}
 In a sedimentation profile, the steady state is flux-free leading to a vanishing mean current $\vec J\equiv \langle \hat {\vec J} \rangle=0$, so that
\begin{equation}\label{eq:bilanMT}
v_s \rho(\vec r) \vec e_z = \sum_i \langle v_0 \vec u(\theta_i) \delta(\vec r-\vec r_i)\rangle 
\end{equation}
where the angular brackets are steady-state averages and we have introduced $\rho\equiv \langle \hat \rho\rangle$. Equation~\eqref{eq:bilanMT} simply states that the downward contribution to the density current due to the sedimentation of the particles is opposed by an average upward bias of their  active force orientations as already shown in the study of the polarization in  section \ref{sec:sedim}. For non-interacting active Brownian particles, the r.h.s. of Eq.~\eqref{eq:bilanMT} can be rewritten as the flux of active impulse (See~\cite{Fily2018JPA} and~\ref{AppendSed})
\begin{eqnarray}
 \langle \sum_i v_0 \vec u(\theta_i) \delta(\vec r - \vec r_i) \rangle &=& - \nabla \cdot \langle \sum_i \dot {\vec r_i} \mu \Delta \vec p^{\rm a}_i  \delta(\vec r - \vec r_i) \rangle \nonumber\;.
 \end{eqnarray}
Integrating Eq.~\eqref{eq:bilanMT} from $z$ to $\infty$, projecting along $\vec e_z$, and dividing by $\mu$ then leads to
\begin{equation}~\label{eq:actimpsedimMT}
\Pi_{\rm w}(z) = \int_z^\infty m g_{\rm eff} \rho(z) =  \langle \sum_i  \dot { z_i} \Delta p^{\rm a}_{z,i} \delta(\vec r - \vec r_i) \rangle \equiv \Pi_{\rm m}(z)\;.
\end{equation}
where $\Pi_{\rm m}(z)$ is the upward flux of active impulse. Note that, here and thereafter, we retain the $x$-dependence in $\delta(\vec r-\vec r_i)$ for dimensionality reasons, even though the result solely depends on $z$. 
Equation~\eqref{eq:actimpsedimMT} balances the total force exerted on the system above $z$, given by the l.h.s., with the flux of active impulse through the horizontal plane at height $z$. {If we were to include a diffusive contribution to the dynamics~\eqref{eq:LangevinABP}, this balance would become
\begin{equation}~\label{eq:actimpsedimdiff}
 \int_z^\infty m g_{\rm eff} \rho(z) =  \langle \sum_i  \dot { z_i} \Delta p^{\rm a}_{z,i} \delta(\vec r - \vec r_i) \rangle\nonumber + \rho k T - D \partial_z \langle \sum_i  \Delta p^{\rm a}_{z,i} \delta(\vec r - \vec r_i) \rangle
\end{equation}
where the last two terms come from the diffusive fluxes of momentum and active impulse, respectively.} The central result~\eqref{eq:actimpsedimMT} shows that, rather surprisingly in this momentum non-conserving system, the total force density applied above a height $z$, $\Pi_{\rm w}(z)$, is balanced, as in equilibrium, by an upward effective momentum flux, $\Pi_{\rm m}(z)$.

To make connection with our experimental system, we note that, for ABPs, the active impulse~\eqref{eq:activeimpulseABP} can be readily computed as \cite{Fily2018JPA}
\begin{equation}
\Delta \vec p^{\rm a}=\frac{f_p}{D_r} \vec u(\theta(t))\;.
\end{equation}
Rewriting~\eqref{eq:actimpsedimMT} then leads to
\begin{equation}\label{eq:balance2MT}
 \Pi_{\rm m}(z)=\frac{v_0^2}{\mu D_r} \langle\sum_i  \cos^2(\theta_i) \delta(\vec r-\vec r_i) \rangle\nonumber-\frac{v_0 v_s}{\mu D_r}\langle\sum_i  \cos(\theta_i) \delta(\vec r-\vec r_i) \rangle \;.
\end{equation}
Introducing the orientation and nematic fields
\begin{eqnarray}
m_z(\vec r)&=&\langle\sum_i  \cos(\theta_i) \delta(\vec r-\vec r_i) \rangle\label{eq:mz}\\
Q_{zz}(\vec r)&=&\langle\sum_i  \cos(2\theta_i) \delta(\vec r-\vec r_i) \rangle\label{eq:Qzz}\;,
\end{eqnarray}
which solely depend on $z$ in the steady state, Eq.~\eqref{eq:balance2MT} can be rewritten as
\begin{equation}\label{eq:pisfinalMT}
\Pi_{\rm m}(z)= \frac{v_0^2}{2 \mu D_r}[ \rho(z)+Q_{zz}(z)]  -\frac{v_0 v_s}{\mu D_r} m_z(z) \;.
\end{equation}

Equation~\eqref{eq:pisfinalMT} shows that, unlike in equilibrium,
$\Pi_{\rm m}(z)$ measured in a sedimentation profile, and thus $\Pi_{\rm w}(z)$, \textit{do not} give access to the pressure of a bulk homogeneous system of density $\rho_0=\rho(z)$. Indeed, the latter would be given by ${\cal G}=\frac{v_0^2}{2 \mu D_r} \rho_0$. The difference is due to the non-isotropic orientation of the active particles in the sedimentation profile. Note that the local density approximation of the equilibrium case is here generalized into a new local approximation involving $m_z$ and $Q_{zz}$, and not only $\rho(z)$. The fact that these two new fields are different in sedimenting and homogeneous isotropic active systems is the reason why the EOS of the latter {cannot} be directly read from sedimentation experiments; it can however be reconstructed from the joint measurement of $\Pi_{\rm w}$, $m_z$ and $Q_{zz}$ (provided $v_0$, $v_s$, $\mu$ and $D_r$ are known). Note that the difference between $\Pi_{\rm m}(z)$ and ${\cal G}$ vanishes in the limit $v_s/v_0\to 0$, as expected in this effective equilibrium regime in which active particles become indistinguishable from passive colloids with an effective temperature $k T_{\rm eff}=v_0^2/(2\mu D_r)$~\cite{Solon2015EPJST,Palacci2010}.

Using $f(\theta)$ computed in section \ref{sec:sedim}, one can rewrite $m_z$ and $Q_{zz}$ as
\begin{equation}
\label{eq:mzQz}
m_z(z)=\rho(z) \frac{v_s}{v_0} + {o}\Big(\frac{v_s}{v}\Big);\qquad Q_{zz}(z) = \rho(z) \frac{v_s^2}{4 v_0^2} + {o}\Big(\Big(\frac{v_s}{v}\Big)^2\Big)
\end{equation}
so that $\Pi_{\rm m}(z)$ can be rewritten as:
\begin{equation}\label{eq:pi21}
\Pi_{\rm m}(z) = \frac{ v^2_0}{2 \mu D_r} \rho(z) \left(
1-\frac{ 7 v_s^2}{4 v_0^2}+ {o}\Big(\Big(\frac{v_s}{v}\Big)^2\Big)
\right)\;.
\end{equation}
For the experimental system described in section \ref{sec:manip} we measure :
\begin{equation}\label{eq:P_experimental}
\mu \Pi_{\rm w}(z) = v_s \int_z^\textrm{L}  dz\rho(z) + \mu\Pi_{\mathrm{out}}
\end{equation}
where $\textrm{L}$ corresponds to the top of the experimental observation box and
\begin{equation}\label{eq:P_experimentalout}
\mu \Pi_\mathrm{out}=  v_s \int_\textrm{L}^\infty  dz\rho(z).
\end{equation}
We extract $\mu \Pi_\mathrm{out}$ by plotting Eq.~\eqref{eq:P_experimental} against Eq.~\eqref{eq:pi21} for all accessible values of $z$ (See~\ref{Append:Piout}). We could equivalently use Eq.~\eqref{eq:pisfinalMT} but the measurements of $Q_{zz}(z)$ and $m_z(z)$ are noisier and the corresponding estimate of $\Pi_{\rm out}$ slightly less reliable. As shown in Fig.~\ref{fig:PI}, the agreement between the  experimental measurements of $\Pi_{\rm w}(z)$ and our theoretical predictions for $\Pi_{\rm m}(z)$ is very good. This agreement is quantified in Fig.~\ref{fig:PI_19} using a parametric plot. Finally, note that, in our experiments, the ratio $v_s/v_0$ is large enough to lead to a clear polarization of the sedimentation profile, as shown in section~\ref{sec:sedim}. The correction to pressure due to polarization however scales as $(v_s/v_0)^2$ so that, although measurable, the impact of polarization on the pressure measurement is limited (see Fig.~\ref{fig:PI_19}).

\subsection{Connection to mechanical forces}
\label{sec:mechanical_forces}

In equilibrium, the relation between momentum flux and force densities exerted on confining interfaces is well understood. In particular the mechanical pressure, defined as the force density on a confining vessel, is equal to the hydrodynamic pressure defined from a bulk stress tensor so that the former satisfies an equation of state: it does not depend on the details of the confining potential. In active matter, on the other hand, the mechanical pressure is generically not given by an equation of state, except in exceptional cases to which our ABP model~\eqref{eq:LangevinABP} belongs. It is thus natural to wonder whether these results, obtained for bulk homogeneous systems, extend to the case of sedimenting profiles.

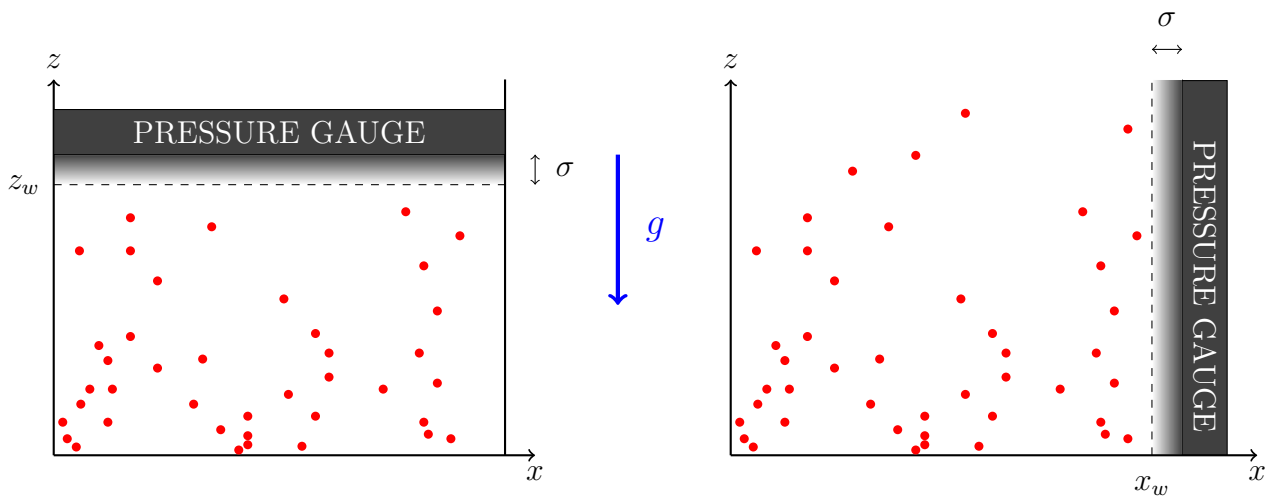
\begin{figure}
  \begin{tikzpicture}[scale=2]
    \shadedraw[white,top color=darkgray, bottom color=white] (0,1.8) rectangle (3,2);
    \draw[thick,->] (0,0) -- (3.2,0) node[below] {$x$};
    \draw[thick,->] (0,0) -- (0,2.5) node[above] {$z$};
    \draw[thick] (3,0) -- (3,2.5);
    \draw[fill=darkgray](0,2) rectangle (3,2.3);
    \draw[ultra thick, white] (1.5,2.15) node {PRESSURE GAUGE};
    \draw (-.2,1.8) node {$z_w$};
    \draw[dashed] (0,1.8) -- (3,1.8);
    \draw[<->] (3.2,1.8) -- (3.2,2);
    \draw (3.4,1.9) node {$\sigma$};

    \fill[red] (.12*3,.11*2) circle(.03);
    \fill[red] (.13*3,.22*2) circle(.03) ;
    \fill[red] (.17*3,.68*2) circle(.03);
    \fill[red] (.33*3,.32*2) circle(.03);
    \fill[red] (.35*3,.76*2) circle(.03);
    \fill[red] (.51*3,.52*2) circle(.03);
    \fill[red] (.55*3,.03*2) circle(.03) ;
    \fill[red] (.58*3,.13*2) circle(.03);
    \fill[red] (.61*3,.26*2) circle(.03);
    \fill[red] (.61*3,.34*2) circle(.03);
    \fill[red] (.73*3,.22*2) circle(.03);
    \fill[red] (.78*3,.81*2) circle(.03);
    \fill[red] (.83*3,.07*2) circle(.03);
    \fill[red] (.81*3,.34*2) circle(.03);
    \fill[red] (.85*3,.48*2) circle(.03);
    \fill[red] (.43*3,.13*2) circle(.03);
    \fill[red] (.23*3,.58*2) circle(.03);
    \fill[red] (.17*3,.79*2) circle(.03);
    \fill[red] (.82*3,.63*2) circle(.03);
    \fill[red] (.90*3,.73*2) circle(.03);

    \fill[red] (.02*3,.11*2) circle(.03);
    \fill[red] (.08*3,.22*2) circle(.03) ;
    \fill[red] (.057*3,.68*2) circle(.03);
    \fill[red] (.82*3,.22*1) circle(.03);
    \fill[red] (.58*3,.81*1) circle(.03);
    \fill[red] (.43*3,.07*1) circle(.03);
    \fill[red] (.31*3,.34*1) circle(.03);
    \fill[red] (.85*3,.48*1) circle(.03);
    \fill[red] (.43*3,.13*1) circle(.03);
    \fill[red] (.23*3,.58*1) circle(.03);
    \fill[red] (.17*3,.79*1) circle(.03);
    \fill[red] (.12*3,.63*1) circle(.03);
    \fill[red] (.10*3,.73*1) circle(.03);

    \fill[red] (.05*3,.11*.5) circle(.03);
    \fill[red] (.03*3,.22*.5) circle(.03) ;
    \fill[red] (.06*3,.68*.5) circle(.03);
    \fill[red] (.88*3,.22*.5) circle(.03);
    \fill[red] (.52*3,.81*.5) circle(.03);
    \fill[red] (.41*3,.07*.5) circle(.03);
    \fill[red] (.37*3,.34*.5) circle(.03);

    \draw[ultra thick, blue,->] (3.75,2) -- (3.75,1);
    \draw[blue] (4,1.5) node {\large $g$};
    
    \begin{scope}[xshift=4.5cm]
    \shadedraw[white,right color=darkgray,  left color=white] (2.8,0) rectangle (3,2.5);
    \draw[fill=darkgray](3,0) rectangle (3.3,2.4925);
    \draw[thick,->] (0,0) -- (3.5,0) node[below] {$x$};
    \draw[thick,->] (0,0) -- (0,2.5) node[above] {$z$};
    \draw[ultra thick, white] (3.15,1.125) node[rotate=270] {PRESSURE GAUGE};
    \draw (2.8,-.2) node {$x_w$};
    \draw[dashed] (2.8,0) -- (2.8,2.5);
    \draw[<->] (2.8,2.7) -- (3,2.7);
    \draw (2.9,2.9) node {$\sigma$};

    \fill[red] (.12*3,.11*2) circle(.03);
    \fill[red] (.13*3,.22*2) circle(.03) ;
    \fill[red] (.17*3,.68*2) circle(.03);
    \fill[red] (.33*3,.32*2) circle(.03);
    \fill[red] (.35*3,.76*2) circle(.03);
    \fill[red] (.51*3,.52*2) circle(.03);
    \fill[red] (.55*3,.03*2) circle(.03) ;
    \fill[red] (.58*3,.13*2) circle(.03);
    \fill[red] (.61*3,.26*2) circle(.03);
    \fill[red] (.61*3,.34*2) circle(.03);
    \fill[red] (.73*3,.22*2) circle(.03);
    \fill[red] (.78*3,.81*2) circle(.03);
    \fill[red] (.83*3,.07*2) circle(.03);
    \fill[red] (.81*3,.34*2) circle(.03);
    \fill[red] (.85*3,.48*2) circle(.03);
    \fill[red] (.43*3,.13*2) circle(.03);
    \fill[red] (.23*3,.58*2) circle(.03);
    \fill[red] (.17*3,.79*2) circle(.03);
    \fill[red] (.82*3,.63*2) circle(.03);
    \fill[red] (.90*3,.73*2) circle(.03);

    \fill[red] (.02*3,.11*2) circle(.03);
    \fill[red] (.08*3,.22*2) circle(.03) ;
    \fill[red] (.057*3,.68*2) circle(.03);
    \fill[red] (.82*3,.22*1) circle(.03);
    \fill[red] (.58*3,.81*1) circle(.03);
    \fill[red] (.43*3,.07*1) circle(.03);
    \fill[red] (.31*3,.34*1) circle(.03);
    \fill[red] (.85*3,.48*1) circle(.03);
    \fill[red] (.43*3,.13*1) circle(.03);
    \fill[red] (.23*3,.58*1) circle(.03);
    \fill[red] (.17*3,.79*1) circle(.03);
    \fill[red] (.12*3,.63*1) circle(.03);
    \fill[red] (.10*3,.73*1) circle(.03);

    \fill[red] (.05*3,.11*.5) circle(.03);
    \fill[red] (.03*3,.22*.5) circle(.03) ;
    \fill[red] (.06*3,.68*.5) circle(.03);
    \fill[red] (.88*3,.22*.5) circle(.03);
    \fill[red] (.52*3,.81*.5) circle(.03);
    \fill[red] (.41*3,.07*.5) circle(.03);
    \fill[red] (.37*3,.34*.5) circle(.03);

    \fill[red] (.88*3,.62*3.5) circle(.03);
    \fill[red] (.52*3,.65*3.5) circle(.03);
    \fill[red] (.41*3,.57*3.5) circle(.03);
    \fill[red] (.27*3,.54*3.5) circle(.03);
    \end{scope}
  \end{tikzpicture}
\caption{Illustration of the measurements of the mechanical pressure
  defined as a force density exerted on a pressure gauge. {\bf Left:}
  Upon inserting a horizontal pressure gauge, the particles are
  confined from above. They start experiencing a repulsive potential
  when reaching a height $z_w$. {\bf Right:} The pressure gauge is now
  oriented along gravity. Particles start experiencing a repulsive
  potential when they reach $x_w$. In both cases, the repulsive
  potential has a finite range $\sigma$ so that no particle goes beyond
  $z_w+\sigma$ or $x_w+\sigma$. }\label{fig:walls}
\end{figure}

To answer this question, we compare the `bulk' pressures $\Pi_{\rm w}(z)$, defined in~Eq. \eqref{eq:psedim}, and its expression as an effective momentum flux, $\Pi_{\rm m}$, computed as Eq.~\eqref{eq:pisfinalMT}, to the mechanical pressure felt by (semi-permeable) pressure gauges. We first consider a pressure gauge modelled as a confining potential starting at height $z_w$, invariant by translation along $x$, and confining the particles from above (See Fig.~\ref{fig:walls}). We thus measure a force density along the $\hat z$ axis, that we call $P_z$. We then turn to the complementary problem of a vertical confining potential starting at $x_w$ and compute the corresponding force density $P_x$ (See Fig.~\ref{fig:walls}). 

Taking into account the confining potential $V_{w}$, the dynamics of the system becomes
\begin{equation}\label{eq:dynpot}
\dot {\vec r}= v_0 \vec u(\theta) - v_s \vec e_z  - \mu\nabla V_w(\vec r)\;.
\end{equation}
The mechanical pressure exerted by the particles on the confining gauge boundary can be computed as
\begin{equation}~\label{eq:pressmech}\label{eq:Pizz}
P_z(z_w)=\int_{z_b}^{\infty} \rho(z) \partial_z V_w(z) dz
\end{equation}
where $z_b$ is any height in the bulk of the system, with $z_b \leq
z_w$. Note that the formula~\eqref{eq:pressmech} is completely generic
and holds for any confining potential. It does not depend on the
choice of $z_b$ since $V_w$ vanishes for $z\leq z_w$. Furthermore, the
convergence of the integral as $z\to\infty$ is ensured by the fact
that the density of particle vanishes in the wall.

Using standard methods~\cite{Solon2015natphys} detailed
in~\ref{app:force}, one gets that
\begin{equation}\label{eq:prisedeteteMT}
P_{z}(z_w)
+W(z_w)= \frac{v_0^2}{2 \mu D_r}[\rho(z_w)+Q_{zz}(z_w)]-\frac{v_0 v_s}{\mu D_r} m_z(z_w)\;,
\end{equation}
where we have introduced the weight $W(z)=\int_{z}^\infty m g_{\rm eff} \rho(z) dz $ of active particles above a given height $z$ \textit{in the presence of the confining boundary}.

Crucially, comparing~\eqref{eq:pisfinalMT} with~\eqref{eq:prisedeteteMT} shows that measuring the weight $\Pi_{\rm w}(z_w)$ of active particles above a given height $z_w$ in an unbounded sedimentation profile (or the effective momentum flux $\Pi_{\rm m}(z_w)$ at this height) \textit{is not} equal to the force density $P_{z}(z_w)$ felt by the pressure gauge. This has two origins that we now detail. First, the r.h.s. of~\eqref{eq:prisedeteteMT}, even though functionally identical to the r.h.s. of~\eqref{eq:pisfinalMT}, will take a different value because of the presence of the confining interface: $\rho(z_w)$, $m_z(z_w)$ and $Q_{zz}(z_w)$ take different values with or without the confining potential. More importantly, the factor $W(z_w)$ tells us that the force that the wall has to exert to confine active particle is equal to their effective momentum flux \textit{minus} the force exerted by the gravity field on the active particles in the wall region. The mechanical pressure $P_{z}(z_w)$ is thus lower than $\Pi_{\rm m}(z_w)$ (even measured in the presence of the confining interface) because part of the confinement is done by gravity itself. This contribution is clearly wall-dependent and will thus always prevent the existence of an equation of state for $P_{z}(z_w)$. 

Interestingly, the computation of the mechanical pressure felt by a confining interface for, say, an equilibrium ideal gas would lead to $P_{z}(z_w)+W(z_w)=\rho(z_w) k T$. Strictly speaking, there is thus no equation of state for the mechanical pressure $P_{z}$ measured in a sedimentation profile of an equilibrium systems. Note that this does not contradict the statistical mechanics definition of pressure $P=-\frac{\partial F}{\partial V}$: this only leads to boundary-independent equation of states when the boundary contributions to the free energy are negligible. This only applies in systems in which the free energy is extensive, which is not the case for sedimenting systems. That said, the weight (density) of passive particles interacting with the pressure gauge is of the order of $\rho m g \sigma$, where $\sigma$ is the interaction range of the confining potential; for non-interacting sedimenting particles, it has to be compared with a pressure $\rho kT$ so that the violation of the equation of state is measured by $\sigma m g / k T$. This is the ratio between the range of the confining potential and the sedimentation height; for sedimenting colloidal particles, this would be completely negligible. On the contrary, for an active system, there is a finite boundary layer of particles that accumulate at the wall, which makes this contribution non-vanishing even when $\sigma \to 0$. Using standard results on the accumulation of active particles at confining boundaries~\cite{Elgeti2015}, we find this contribution to be of the order of $v_s/v_0$\footnote{The bulk pressure indeed scales as $\rho v_0^2/(2 \mu D_r)$ while the weight of particles at the boundary scales as $(v_s/\mu) \rho v_0/D_r$}. 

Let us now look at what happens for a confining boundary oriented normal to the $\hat x$ direction (See Fig.~\ref{fig:walls}). The exact same computation as above leads to
\begin{eqnarray}\label{eq:pressuremech}
P_x(z)&=& kT \rho(x_b,z) + \frac{v_0^2}{2 \mu D_r}[\rho(x_b,z)+Q_{xx}(x_b,z)]\;.
\end{eqnarray}
where $(x_b,z)$ corresponds to a point in the bulk of the system. Interestingly, this $z$-dependent force density satisfies an equation of state, and will not depend on the details of the confining boundary. It is entirely predicted by bulk properties of the fluid, but, again, these properties are not measured by $\Pi_{\rm w}(z)$, since~\eqref{eq:pressuremech} differs from~\eqref{eq:pisfinalMT} because the fluid is anisotropic.
Note that for weakly sedimenting systems, however, Eq.~\eqref{eq:pi21} shows the difference between $P_x(z)$ and $\Pi_{\rm w}(z)$ to vanish so that a pressure gauge could be used, in principle, to access $\Pi_{\rm w}(z)$ or $\Pi_{\rm m}(z)$ when the self-propulsion speed is much larger than the sedimentation velocity.

\section*{Conclusion}

In this paper we have used active Janus colloids and active Brownian particles to study sedimenting active particles. We have shown experimentally that when the sedimentation speed is comparable to the propulsion speed, a net polarization of the active colloids develops in the bulk of the system. The theoretical predictions for
the distribution of orientations of the particles agree very well with the experimental results without any free parameter. We then discussed different definitions of pressure for sedimenting active particles. Using our ABP model, we have shown that the bulk pressure defined as the weight exerted on the system above a certain height can be interpreted in terms of local momentum transfer. This is verified to a very good approximation in our experiments. Finally we discussed when such bulk definitions of pressure can be related to local force densities exerted on pressure gauges.

\appendix

\section{Measuring $v_s$ and maximum $v_s/v_0$ ratio}
\label{sec:vs}

\begin{figure}
	\centering
  	\includegraphics[width=0.6\columnwidth]{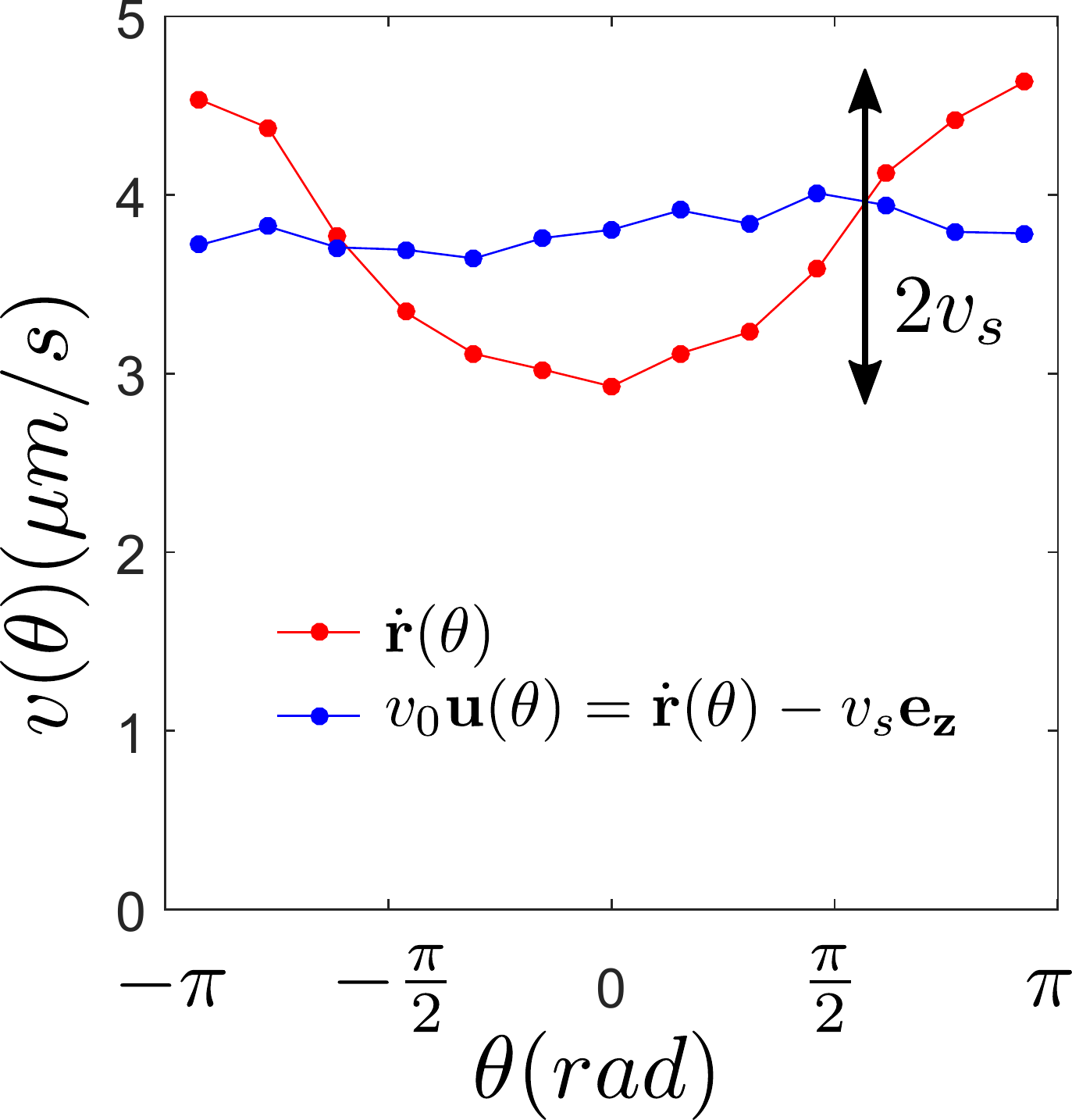}
	\caption{{\bf Red}, $\dot \vec r (\theta)$ is not isotropic due to $v_s$ contribution. We measure $v_s$ using $\dot \vec r (0)-\dot \vec r (\pi)=2v_s$. {\bf Blue}, as expected $v_0\vec u(\theta)= \dot {\vec r}-v_s \vec e_z$ becomes isotropic when we remove $v_s$ contribution. }
	\label{fig:Vs}
\end{figure}

\paragraph*{$v_s$ measurement.}
Getting a precise measure of $v_s$ is essential as we do not have direct access to $v_0 \vec u(\theta)$ but to $\dot {\vec r}=v_0\vec u(\theta) -v_s \vec e_z$.
While we measure the tilt of the experimental chamber with the resulting $\alpha$ angle, and could $a~priori$ compute $v_s$, the resulting error bars are too high for a quantitative study.
We thus decided to measure it $a~posteriori$ using the 2D probability density function, and the fact that $\dot r (0)-\dot r (\pi)=2v_s$.
This gives much more precise measurement of $v_s$ and indirectly $v_0$ and $f(\theta)$.
We verify in Fig.~\ref{fig:Vs} that $\dot r (\theta)$ is not isotropic due to $v_s$ contribution (red), but that as expected $v_0\vec u(\theta)= \dot {\vec r}-v_s \vec e_z$ becomes isotropic when we remove $v_s$ contribution (blue).

\paragraph*{Maximum $v_s/v_0$ ratio.}
In the experiments the ratio $v_s/v_0$ varies from $\sim 0.09$ to $\sim 0.29$. Higher ratios are experimentally difficult to access quantitatively as the gas phase is so sparse that there is not enough consistent data. 
Another difficulty arises from the fact that we are tilting solely the experimental chamber, and not the full microscope. This induces strong defocus for the highest $v_s/v_0$ ratio, which limits the effective size of observation.
While it would be possible to tilt the full experiment, including the microscope, the control and precision would then be much more difficult than with a piezoelectric device.

\section{Projected 3D}
\label{sec:3dproj}

In the main text, we modeled the experimental Janus colloids as ABPs living in the two dimensions of the bottom plate. However, in practice, the orientation of the colloids is not constrained to the 2D plane and can venture in the third dimension. We show in this Appendix that taking into account the 3D orientation of the particles leads to very small corrections to the predictions of the 2D model, so that the two cannot be distinguished by the experiments.

In 3D, the Active Brownian dynamics of Eq.\ref{eq:LangevinABP} now read
\begin{equation}
  \label{eq:LangevinABP3D}
  \dot {\vec r}=v_0\vec u-v_s \vec e_z; \qquad \dot{\vec u}=\vec u\times\sqrt{2D_r}\Gvec\xi
\end{equation}
where $\Gvec\xi$ is a 3D vector of Gaussian white noises with zero mean and unit variance $\langle \xi_i(t)\xi_j(t')\rangle = \delta_{ij}\delta(t-t')$, $i$ and $j$ being Cartesian coordinates. (Eq.~\eqref{eq:LangevinABP3D} uses Stratonovich convention.)
Contrary to ABPs in 2D, we do not have an exact solution for this model and thus resort to simulations.

\begin{figure}
	\centering
  	\includegraphics[width=0.6\columnwidth]{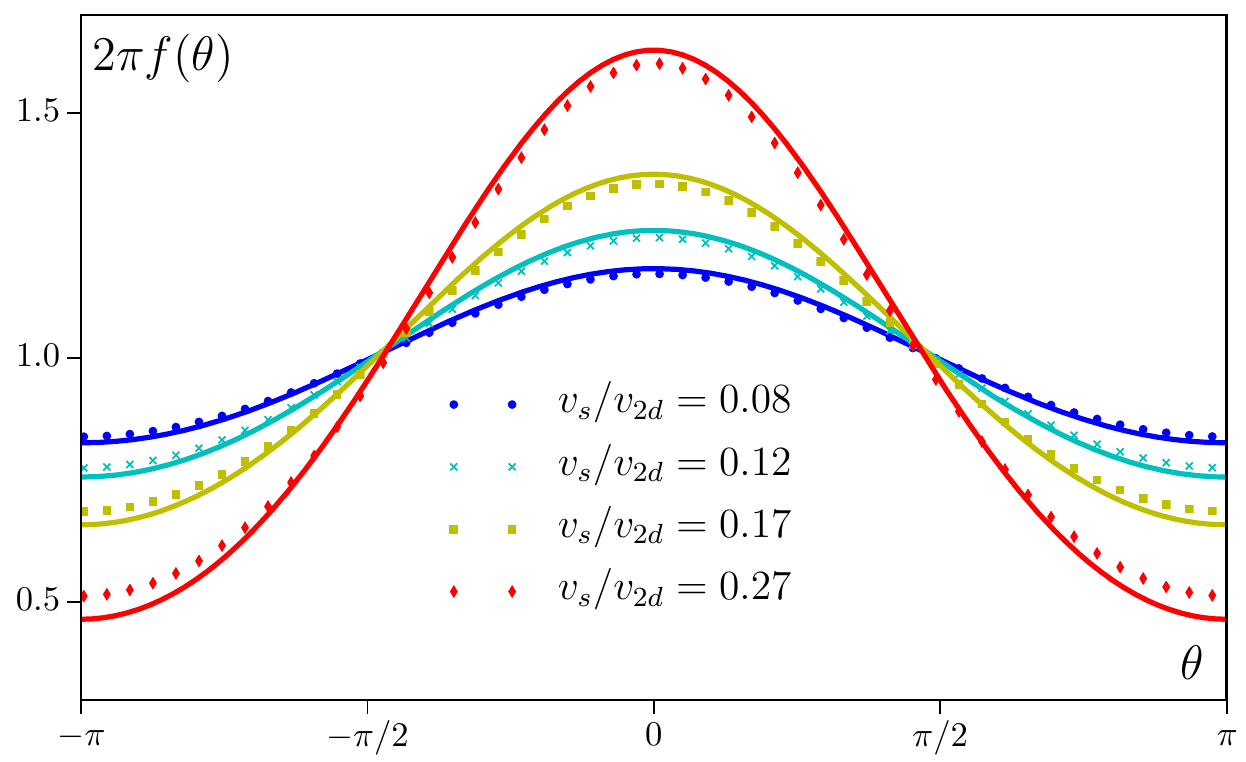}
	\caption{Exact angular distribution for 2D ABPs (plain lines) and angular distribution for 3D ABPs measured in simulations of Eq.\ref{eq:LangevinABP3D} (symbols).}
	\label{fig:ftheta_2Dvs3D}
\end{figure}

The orientation $\vec u$ can be parametrized in 3D by two angles $\theta$ and $\phi$. We take $\theta$ as before in the (x-z)-plane with $\theta=0$ along the $z$-axis so that, once integrated over $\phi$ we can compare the angular distributions $f_{3d}(\theta)$ measured in simulations of Eq.~\ref{eq:LangevinABP3D} and the analytical result for $f_{2d}(\theta)$. The two are compared in Fig.\ref{fig:ftheta_2Dvs3D} for the values of $v_s/v_{2d}$ corresponding to the experiments, taking into account that, for the 3D model, the speed $v_0$ appearing in Eq.~\ref{eq:LangevinABP3D} is related to the speed measured in the 2D bottom plane by a geometric factor $v_{2d}=(\pi/4)v_0$. The difference between the distributions predicted by the 2D and 3D models is smaller than 5\% and the experiments thus cannot discriminate between the two.

\section{Measuring $\mu \Pi_\mathrm{out}$}
\label{Append:Piout}

We discuss how to measure in experiments the pressure at height $z$, defined as the weight above this position:
\begin{equation}\label{eq:Pi_base}
\mu \Pi_{\rm w}(z) = v_s \int_z^\infty  dz\rho(z)\;.
\end{equation}

In theory, for an open system, the density profile follows an exponential decay, and should only vanish at $z=+\infty$.
In the experiment, however, we can only integrate the density profile up to a height $\textrm{L}$, because we lose particles that are out of the experimental window, or due to the defocus of the microscope. We call the missing contribution $\Pi_{\mathrm{out}}$ and write
\begin{equation}
\mu \left(\Pi_{\rm w}(z)-\Pi_{\mathrm{out}} \right)= v_s \int_z^\textrm{L}  dz\rho(z)
\label{eq:Poutannexe}
\end{equation}
\begin{equation}
\mu \Pi_\mathrm{out}=  v_s \int_\textrm{L}^\infty  dz\rho(z).
\end{equation}

\begin{figure}
	\centering
  	\includegraphics[width=0.9\columnwidth]{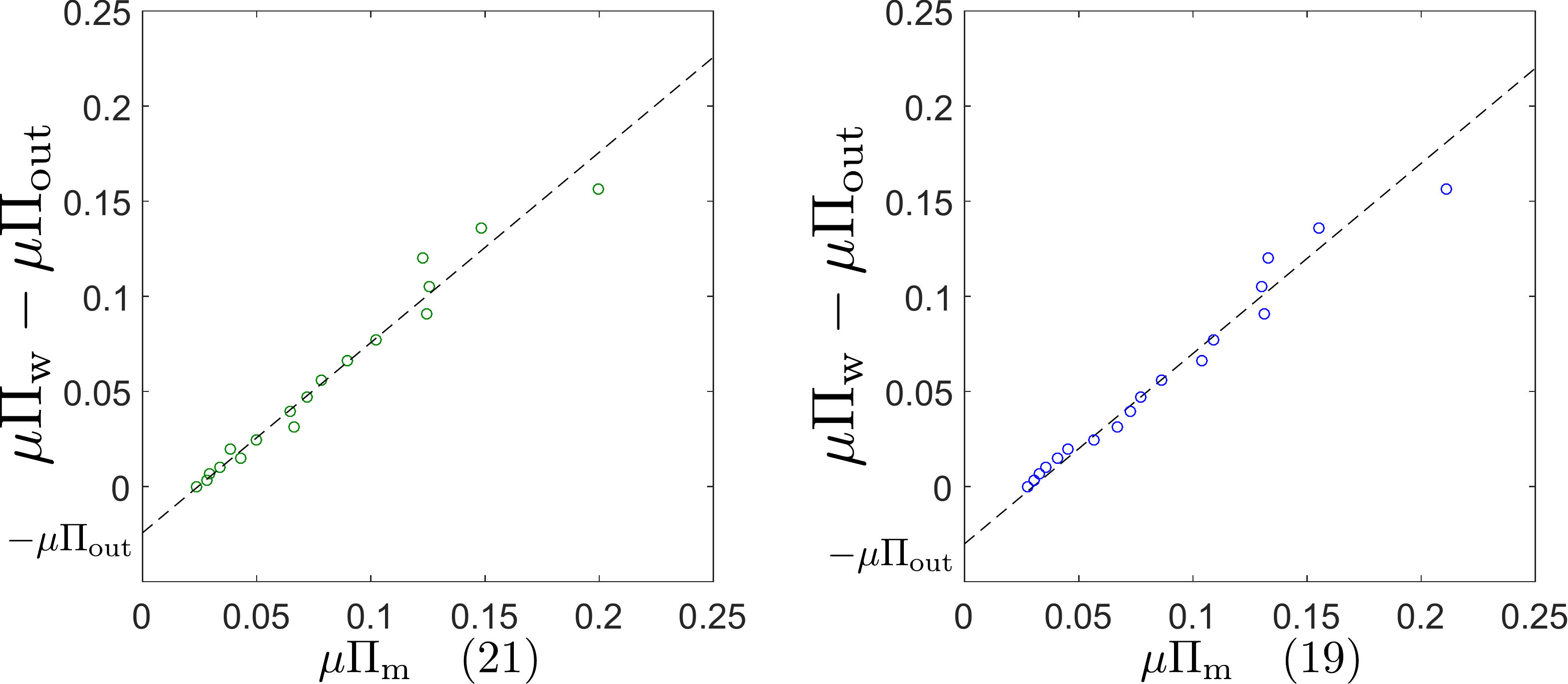}
	\caption{$\mu \Pi_w - \mu\Pi_{\mathrm{out}} = v_s \int_z^\textrm{L}  dz\rho(z)$ vs $\mu \Pi_m$ (21) (left) and (19) (right)  for experiment with $v_s/v_0=0.29$. We use this plot to measure $\mu \Pi_{\mathrm{out}}$ by looking at the intersect between the affine fit $y=x+b$ and the y-axis. We find $\mu \Pi_{\mathrm{out}} = 0.03 \pm 0.005 s^{-1}$ for (19), and $\mu\Pi_{\mathrm{out}} = 0.024 \pm 0.004 s^{-1}$ for (21).}
	\label{fig:Pi_out}
\end{figure}

\begin{figure}
	\centering
  	\includegraphics[width=0.9\columnwidth]{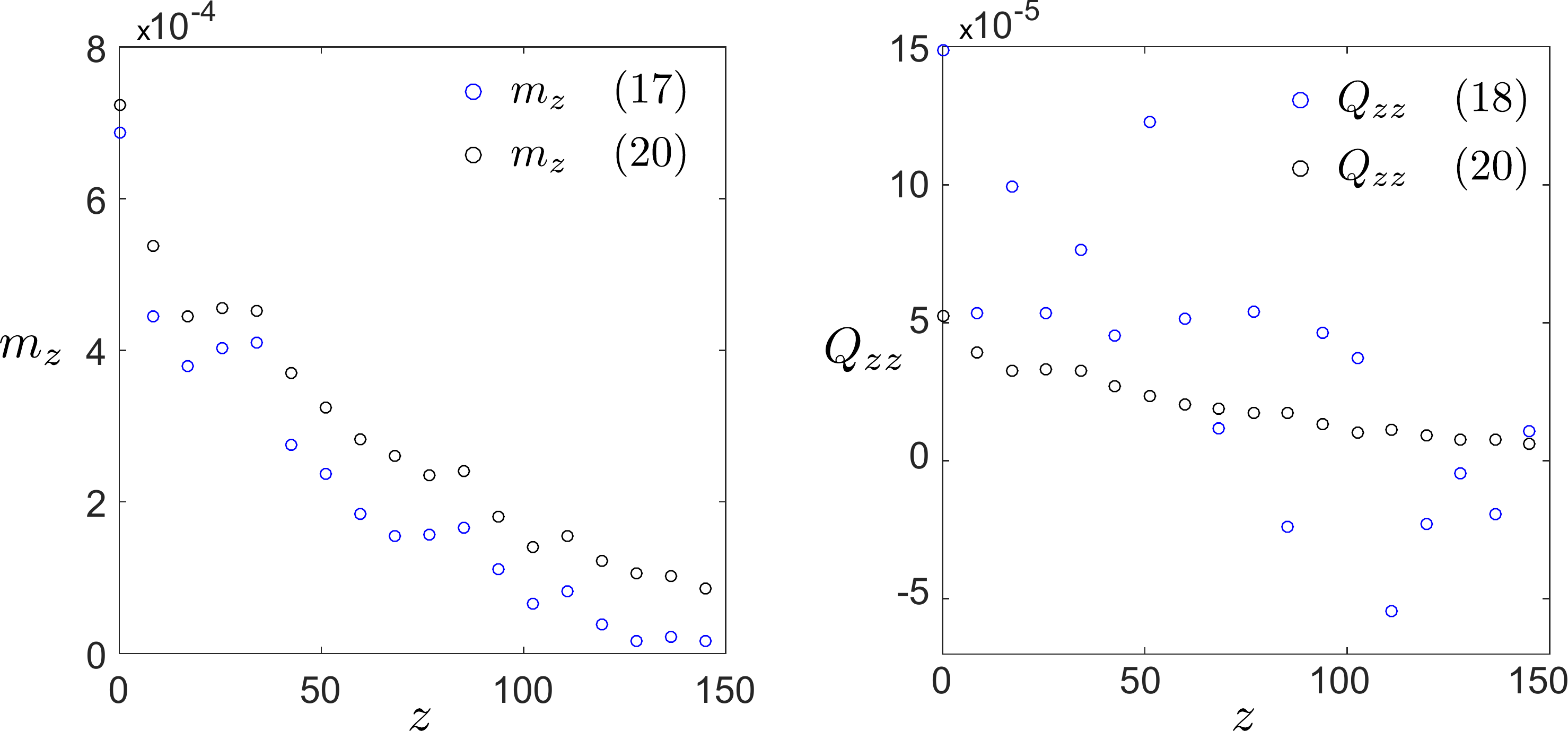}
	\caption{Blue circles: experimental measurements for $m_z$ (left) and $Q_{zz}$ (right) vs $z$, using Eq.~\eqref{eq:mz} and~Eq.\eqref{eq:Qzz}.  Black circles are the theoretical values from Eq.~\eqref{eq:mzQz}.} 
    %Note that the experimental data for $m_z$ are systematically lower than theory, and the experimental data for $Q_{zz}$ are extremely noisy, hence the reason we chose to use Eq.~\eqref{eq:mzQz}.}
	\label{fig:MQ}
\end{figure}

We extract $\mu \Pi_\mathrm{out}$ using a parametric plot of Eq.~\eqref{eq:Poutannexe} against Eq.~\eqref{eq:pi21} for all accessible values of $z$.
We obtain an affine relationship (see~Fig.\ref{fig:Pi_out}, left). We then measure $- \mu \Pi_\mathrm{out}$ at the intersect between the affine fit and the y-axis, and find $\mu \Pi_{\mathrm{out}} = 0.024 \pm 0.004 s^{-1}$.
Note that we could equivalently use Eq.~\eqref{eq:pisfinalMT} (see~Fig.\ref{fig:Pi_out}, right) but the measurements of $Q_{zz}(z)$ and $m_z(z)$ are noisier (Fig.~\ref{fig:MQ}) and the corresponding estimate of $\Pi_{\rm out}=0.03 \pm 0.005 s^{-1}$ slightly less reliable.

\section{Sedimentation and active impulse}
\label{AppendSed}
For sake of generality, we derive in this appendix the relationship between the bulk pressure $\Pi_{\rm w}(z)$, as defined in~\eqref{eq:psedim}, and momentum and active impulse transfers in the bulk of the active system, in the presence of translational diffusion. We consider particles evolving with the dynamics
\begin{equation}
\dot {\vec r}= v_0 \vec u(\theta) - v_s \vec e_z + \sqrt{D} \Gvec\eta\;.
\end{equation}
Using It\=o calculus, the dynamics of the exact microscopic density field $\hat \rho=\sum_i \delta(\vec r- \vec r_i)$ is given by~\cite{solonPressure2015b,Dean1996JPA}
\begin{eqnarray}
\dot {\hat \rho} &=& - \nabla \cdot \hat {\vec J}\\
\hat {\vec J} &\equiv& - D \nabla \hat \rho + \sqrt{2 D \hat \rho} \Lambda - v_s \vec e_z \hat \rho + \sum_i v_0 \vec u(\theta_i) \delta(\vec r -\vec r_i)
\end{eqnarray}
where $\Lambda(\vec r,t)$ is a $\delta$-correlated Gaussian white noise field of zero mean and unit variance. In a sedimentation profile, the steady state is flux-free leading to a vanishing mean current $\vec J\equiv \langle \hat {\vec J} \rangle=0$, so that
\begin{equation}\label{eq:bilan}
v_s \rho(z) = \sum_i \langle v_0 \cos\theta_i \delta(\vec r-\vec r_i)\rangle - D \partial_z \rho(z) 
\end{equation}
where we have introduced $\rho\equiv \langle \hat \rho\rangle$ and we have used that the system is invariant by translation along $\hat {\vec x}$ to replace $\nabla$ by $\partial_z$ and to drop any useless dependence on $x$. Equation~\eqref{eq:bilan} simply states that the downward contribution to the density current due to the sedimentation of the active particles is opposed both by the average upward motion of the particles and by the upward diffusive flux. For non-interacting active Brownian particles, the first term of the r.h.s. of Eq.~\eqref{eq:bilan} can be rewritten as the flux of active impulse~\cite{Fily2018JPA}. Using It\=o calculus, one writes
\begin{eqnarray}
\partial_t \langle \cos\theta_i \delta(\vec r - \vec r_i) \rangle &=& - \nabla \langle \dot {\vec r_i} \cos\theta_i \delta(\vec r - \vec r_i) \rangle+D \Delta \langle \cos\theta_i \delta(\vec r - \vec r_i) \rangle \nonumber\\
&&- D_r \langle \cos\theta_i \delta(\vec r - \vec r_i) \rangle   
\end{eqnarray}
so that, in steady state,
\begin{eqnarray}
 \langle \cos\theta_i \delta(\vec r - \vec r_i) \rangle &=&  \partial_{zz} \langle \frac{D}{D_r} \cos\theta_i \delta(\vec r - \vec r_i) \rangle- \partial_z \langle \frac{\dot {\vec z_i}}{D_r}  \cos\theta_i \delta(\vec r - \vec r_i) \rangle 
\end{eqnarray}
Integrating Eq.~\ref{eq:bilan} from $z$ to $\infty$ then leads to
\begin{eqnarray}~\label{eq:actimpsedim}
v_s \int_z^\infty \rho(z) &=& D \rho(z) + \langle \sum_i \frac{v_0}{D_r} \dot {\vec z_i} \cos\theta_i \delta(\vec r - \vec r_i) \rangle\nonumber\\&& -D\partial_{z} \langle \sum_i\frac{v_0 }{D_r} \cos\theta_i \delta(\vec r - \vec r_i) \rangle\;.
\end{eqnarray}
Writing $v_0=\mu f_p$, with $f_p$ the propulsive force of an active particle, and dividing Eq.~\eqref{eq:actimpsedim} by $\mu$ then leads to the effective momentum balance equation
\begin{eqnarray}~\label{eq:actimpsedim2}
\Pi_{\rm w}(z)&=& k T \rho(z) + \langle \sum_i \dot {z_i} \frac{f_p}{D_r}  \cos\theta_i \delta(\vec r - \vec r_i) \rangle\nonumber\\
&& -D\partial_{z} \langle\sum_i \frac{f_p }{D_r} \cos\theta_i \delta(\vec r - \vec r_i) \rangle\equiv \Pi_{\rm m}(z)
\end{eqnarray}
which is Eq.~\eqref{eq:actimpsedimdiff} of the main text. 

We note that the same result can also be obtained, more in line with~\cite{Fily2018JPA}, by considering an underdamped system and taking the overdamped limit at the end.

\section{Force density exerted by active Brownian particles in a sedimentation profile}
\label{app:force}
In this appendix we detail the computation of the force density exerted by an active system on a confining interface located at height $z_w$:
\begin{equation}
P_{z}(z_w)=\int_{z_b}^{\infty} \rho(\vec r) \partial_z V_w(\vec r) dz\;.
\end{equation}
As before, we compute the dynamics of $\rho(\vec r)$ and find, in the steady state
\begin{equation}
\dot \rho(\vec r)=0=-\partial_z J_z(\vec r)
\end{equation}
where $J_z$ is the mean density current along $\hat z$, which vanishes in the steady state:
\begin{equation}
J(\vec r)=0=-v_s \rho(\vec r) -\rho(\vec r)\mu \partial_z V_w  + v_0 m_z(\vec r) \;.
\end{equation}
This allows us to write the mechanical pressure as
\begin{equation}
P_{z}(z_w)=-\frac{v_s}{\mu}\int_{z_b}^\infty dz \rho(\vec r)  +  \frac{v_0}{\mu}\int_{z_b}^{\infty}m_z(\vec r) dz\;.
\end{equation}
Using It\=o calculus, we now compute the dynamics of $m_z$, defined in~\eqref{eq:mz}, which yields in the steady-state
\begin{equation}
\dot m_z =0 =-\partial_z\langle \sum_i \cos\theta_i \dot z_i \delta(\vec r-\vec r_i)\rangle - D_r m_z \;.
\end{equation}
Together with Eq.~\eqref{eq:dynpot}, one gets 
\begin{equation}
m_z = -\partial_z[v_0\frac{\rho+Q_{zz}}{2 D_r}-\frac{v_s}{D_r} m_z -\frac{\mu}{D_r} m_z \partial_z V_w ]\;.
\end{equation}
Therefore, the mechanical pressure felt by the boundary is given by
\begin{eqnarray}
P_{z}&=&-\frac{v_s}{\mu}\int_{z_b}^\infty dz \rho(\vec r) + \frac{v_0^2}{2 \mu D_r}[\rho(\vec r_b)+Q_{zz}(\vec r_b)] -\frac{v_0 v_s}{\mu D_r} m_z(\vec r_b)
\end{eqnarray}
This expression is valid for any $z_b\leq z_w$. In particular, for $z_b=z_w$ we recover Eq.~\eqref{eq:prisedeteteMT} of the main text.

\end{document}